\documentclass[conference]{IEEEtran}
\IEEEoverridecommandlockouts
\usepackage{cite}
\usepackage{amsmath,amssymb,amsfonts}
\usepackage{algorithmic}
\usepackage{graphicx}
\usepackage{textcomp}
\usepackage{xcolor}
\usepackage{subcaption}
\usepackage{amsmath}
\usepackage{amssymb}
\usepackage{gensymb}
\usepackage{multirow}
\usepackage{array}
\def\BibTeX{{\rm B\kern-.05em{\sc i\kern-.025em b}\kern-.08em
    T\kern-.1667em\lower.7ex\hbox{E}\kern-.125emX}}
\begin{document}

\title{Frequency-responsive RCS characteristics and scaling implications for ISAC development}

\author{Saúl Fenollosa$^+$; Monika Drozdowska$^+$; Wenfei Yang*; Sergio Micó-Rosa$^+$; Alejandro Castilla$^+$; \\
Alejandro Lopez-Escudero$^+$; Jian Li*; Narcis Cardona$^+$ \\
\textit{$^+$iTEAM Research Institute, Universitat Politècnica de València, Spain} \\ 
\textit{*Wireless Technology Laboratory, Huawei Technologies Co., Ltd., China} \\
\textit{$^+$[sjfenarg, mdrozdo, sermiro, a.castilla, alloes3, ncardona]@upv.edu.es} \\
\textit{*[yangwenfei4, calvin.li]@huawei.com}
\vspace{-2pt}
}

\maketitle

\begin{abstract}
This paper presents an investigation on the Radar Cross-Section (RCS) of various targets, with the objective of analysing how RCS properties vary with frequency. Targets such as an Automated Guided Vehicle (AGV), a pedestrian, and a full-scale car were measured in the frequency bands  referred to in industry standards as FR2 and FR3. Measurements were taken in diverse environments, indoors and outdoors, to ensure comprehensive scenario coverage. The methodology employed in RCS extraction performs background subtraction, followed by time-domain gating to isolate the influence of the target. This analysis compares the RCS values and how the points of greatest contribution are distributed across different bands based on the range response of the RCS. Analysis of the results demonstrated how RCS values change with frequency and target shape, providing insights into the electromagnetic behaviour of these targets. Key findings highlight how much scaling RCS values based on frequency and geometry is complex and varies among different types of materials and shapes. These insights are instrumental for advancing sensing systems and enhancing 3GPP channel models, particularly for Integrated Sensing and Communications (ISAC) techniques proposed for 6G standards.
\end{abstract}
\vspace{-2pt}

\section{Introduction} 

In the context of advanced mobile communication networks, Integrated Sensing and Communication (ISAC) technology has emerged as a forefront innovation for future 6G networks. ISAC aims to integrate sensing with communications, enabling network elements to function as sensors for gathering environmental information \cite{ISAC_general}. 
This approach offers a wide range of potential sensing applications, while enhancing spectral efficiency, leveraging radio channel information derived from sensing data \cite{ISAC_applications}.
Existing literature proposes ISAC channel models that combine stochastic standards with deterministic information provided by the detection and characterization of environmental elements \cite{ISAC_channel1}. 
One of the steps to undertake is the analysis of the influence of usual elements in radio channels, such as pedestrians or cars \cite{matsunami2012, schipper2011}. It may be done by analysing the Radar Cross-Section (RCS), which is defined as a metric of the detectability of an object by radar. This metric depends on the element's size and shape, its material, and the frequency at which it is observed. RCS information helps constraining the stochastic information according to 3GPP channel standards, allowing the addition of a deterministic metric to the ISAC channel model  \cite{ISAC_channel2}.

This work outlines the process of measuring RCS using a vector network analyzer (VNA) for three targets at various elevations and across two  frequency bands: 10-15 GHz and 25.75-30.25 GHz. The measured targets include a pedestrian, an Automated Guided Vehicle (AGV), and a full-scale car. Additionally, this work elaborates on the complete process of extracting the influence of the target from other components of the scenario, using techniques such of background subtraction and time-domain gating. It continues with an analysis of the feasibility of performing frequency scaling on complex targets, and ends with an analysis of the spatial points with the greatest influence on the RCS value, comparing their consistency in both bands.

The rest of the paper is organised as follows: Section II describes the RCS extraction methodology, Section III outlines the measurement campaign, Section IV discusses the results, and Section V summarises the conclusions of this paper.

\vspace{-3pt}
\section{RCS Extraction Methodology}

\begin{figure}
\centering
\begin{subfigure}{.34\textwidth}
  \centering
\includegraphics[width=.9\linewidth]{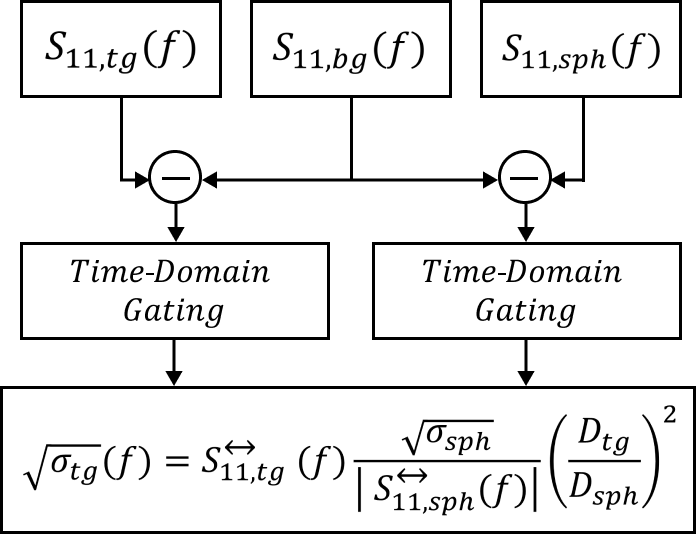}
\end{subfigure}%
\caption{RCS extraction methodology diagram.}
\vspace{-10pt}
\label{fig:RCS_blockDiagram}
\end{figure}

\vspace{-3pt}
The methodology for extracting the RCS, based on existing literature \cite{RCS_proc1},\cite{RCS_proc2}, is illustrated in the block diagram shown in Fig. \ref{fig:RCS_blockDiagram}. This process is conducted using complex S-parameter measurements obtained with a VNA. It requires three different measurements: 
\vspace{-3pt}
\begin{itemize}
    \item $S_{11,\text{tg}}(f)$, which measures a scenario with a target in it
    \item $S_{11,\text{bg}}(f)$, which corresponds to the scenario without the target
    \item $S_{11,\text{sph}}(f)$, which measures a calibration object at  the target location in the scenario
\end{itemize}

\subsection{Background subtraction and time-domain gating}

The background influence is subtracted from both the target and the calibration object measurements to eliminate unwanted components, ensuring that only the signals of interest remain. After this subtraction, a time-domain gating algorithm is applied. This algorithm filters out components not belonging to the object under study by considering only reflections within a specific time range. 

\subsection{Frequency dependent RCS definition and calibration procedure}

\begin{figure}
\centering
\begin{subfigure}{.35\textwidth}
  \centering
\includegraphics[width=.9\linewidth]{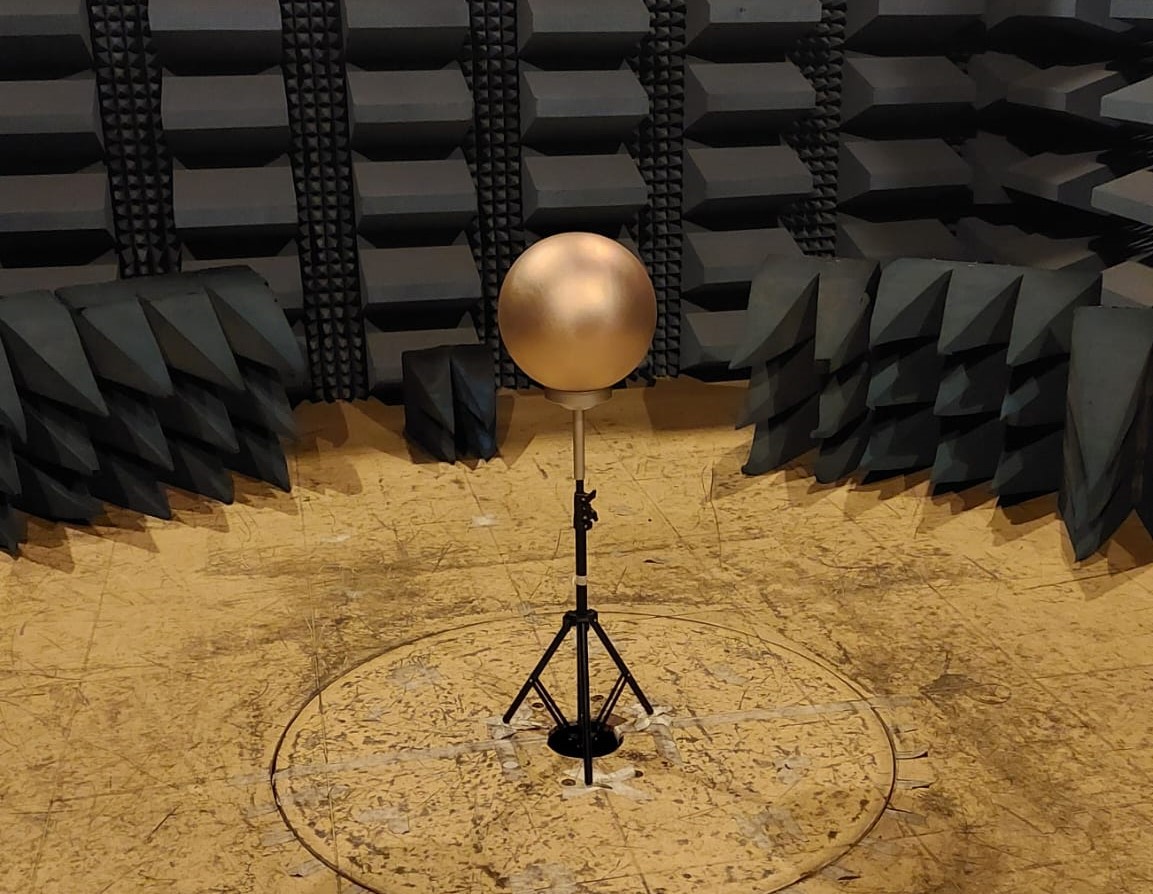}
\end{subfigure}%
\caption{PEC sphere used as calibration object.}
\label{fig:PEC_sphere}
\end{figure}

To enable a more comprehensive analysis of the target's influence using the RCS metric, we define the frequency-dependent root of the RCS, denoted as $\sqrt{\sigma_{tg}}(f)$, whose expression is written in Fig. \ref{fig:RCS_blockDiagram}. This parameter is derived from the signal $S_{11,\text{tg}}^{\leftrightarrow}(f)$, obtained after background subtraction and time-domain gating application to the measured signal with the target's influence. Similarly, the signal $S_{11,\text{sph}}^{\leftrightarrow}(f)$ is obtained for the calibration object, processed in the same manner as for the target. The theoretical square root of the RCS value for the calibration object, denoted as $\sqrt{\sigma_{sph}}$, is then used as a reference. We can accurately calibrate the measurement system by comparing the measured response of the calibration object to its known theoretical RCS. This process compensates for variations in the radiation pattern, antenna gain, and other system losses. The chosen calibration object was a perfect electric conductor (PEC) sphere with a radius of $R = 15$ cm (Fig. \ref{fig:PEC_sphere}). The PEC sphere was selected as the calibration object because of its stable and predictable RCS characteristics, especially its independence from frequency for sufficiently small wavelengths. For that case, the theoretical RCS value is determined solely by the physical dimensions of the sphere and is given by
\begin{equation}
    \sigma_{sph} = \pi R^2,
\end{equation}
where $R$ is the radius of the sphere. Additionally, it is necessary to correct for discrepancies in propagation losses that may arise due to different positions of the target and the calibration object. This correction is achieved by applying a factor derived from the square of the ratio between the distance to the target, $D_{tg}$, and the distance to the calibration object, $D_{sph}$.

The RCS value of the target over the frequency band is obtained by averaging all the squared values of $\sqrt{\sigma_{tg}}(f)$, 
\begin{equation}
\text{RCS} = \frac{1}{N} \sum_{i=1}^{N} \sigma_{tg}(f_i),
\end{equation}
where $N$ is the number of samples measured in the band. However, the metric $\sqrt{\sigma_{tg}}(f)$ enables us to assess the target's characteristics not only in the frequency domain but also in the time and range domains. This is achieved by applying the inverse Fourier transform, expressed as follows:
\begin{equation}
    \sqrt{\sigma_{tg}}(t) = \mathcal{F}^{-1}\left\{\sqrt{\sigma_{tg}}(f)\right\}.
\end{equation}

\section{Measurement campaign}

\subsection{Measurements setup}\label{MS}
The channel sounding system used for the measurements consists of a VNA and one horn antenna connected to the VNA via RF cable. The VNA is the Rohde \& Schwarz ZNB40 \cite{ZNB40} designed to work in the 100~kHz to 40~GHz frequency band. As for the antennas, three different types were utilised depending on the target frequency band. The QSH-SL-10-15-N-20 (QMS-00195) antenna \cite{QMS00195} for 10-14 GHz and PE9855B/SF-15 antenna \cite{PE9} for 10-15 in FR3, and the QWH-SL-18-40-K-SG (QMS-00910) antenna \cite{QWH1840} for 25.75-30.25 GHz in FR2. The RF cable KBL-2M-LOW+ \cite{KBL} was used to connect port 2 of the VNA with the horn antenna. 
Used cables were 2~m long low-loss coaxial cables with an insertion loss of 2.35~dB at 10~GHz and 4.3~dB at 30~GHz. Some additional parameters are summarised in Table~\ref{tb:meas_setup}.

\begin{table}[ht]
	\centering
	\caption{Summary of the measurement parameters.}
	\label{tb:meas_setup}
	\renewcommand{\arraystretch}{1.19}
    \resizebox{0.35\textwidth}{!}{
	\begin{tabular}{|c|c|c|c|}
		\hline
        Center frequency  & 12 & 12.5 GHz & 28 GHz \\
        \hline
        Bandwidth  & 4 GHz & 5 GHz & 4.5 GHz \\
        \hline
        Antenna HPBW  & 16$^\circ$ & 32$^\circ$ & 25$^\circ$ \\
        \hline
		Tx power  & \multicolumn{3}{|c|}{0~dBm}  \\
		\hline
        IF Bandwidth &  \multicolumn{3}{|c|}{100 kHz} \\
		\hline
        No. of samples per CIR &  \multicolumn{3}{|c|}{2001} \\
		\hline
	\end{tabular}}	
\end{table}

\subsection{Measured Targets}\label{MO}
\begin{figure}[htbp]
\begin{subfigure}{.36\textwidth}
\includegraphics[width=\linewidth]{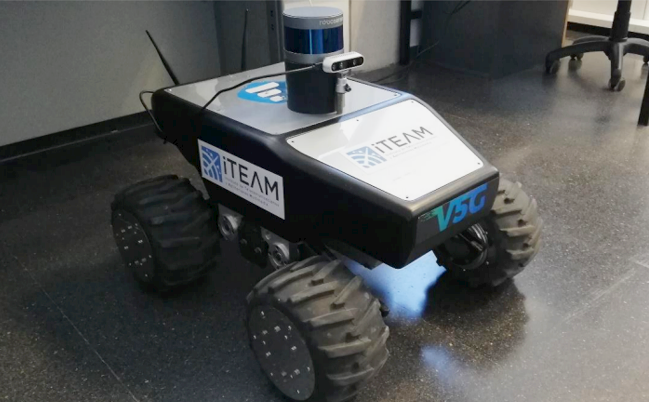}
\centering
\caption{}
\label{fig:agv}
\end{subfigure}%
\hspace{3pt}
\begin{subfigure}{.093\textwidth}
\includegraphics[width=\linewidth]{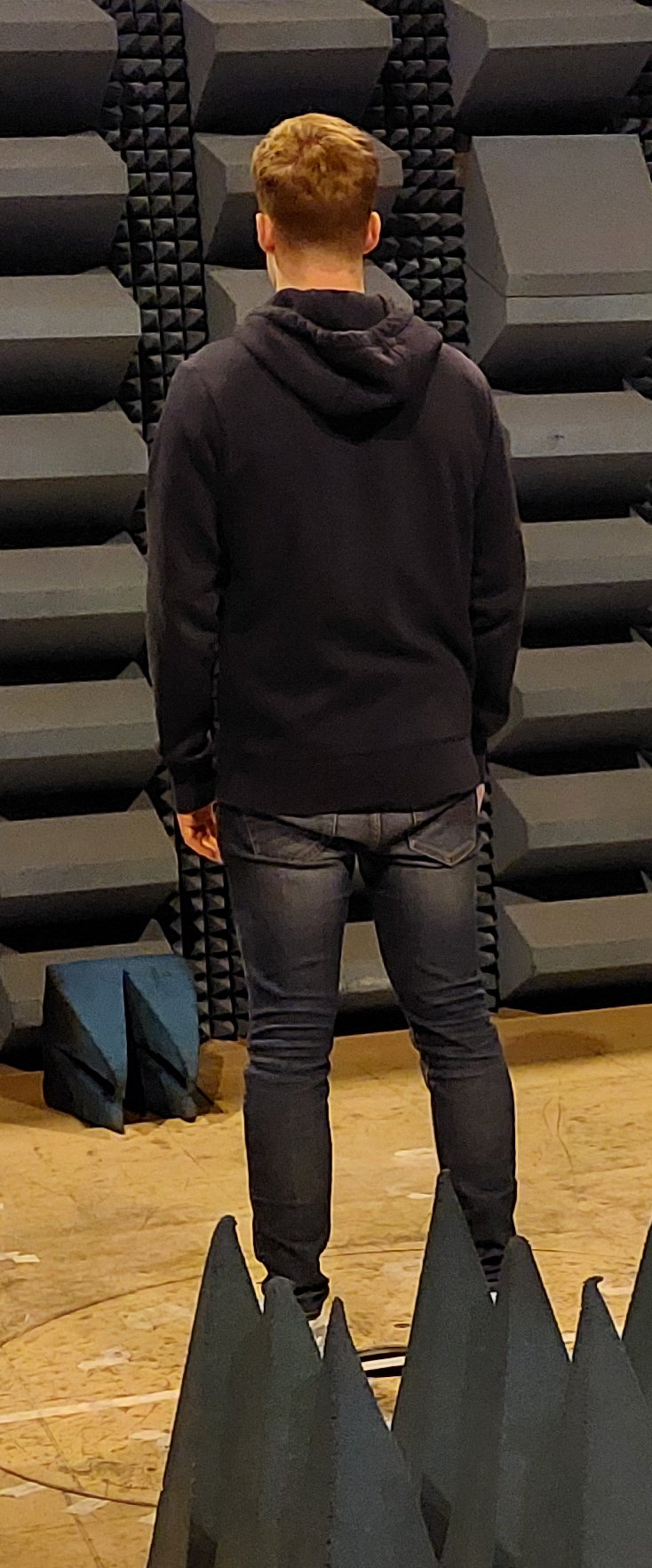}
\centering
\caption{}
\label{fig:uh}
\end{subfigure}%
\caption{Measured targets: a) AGV, b) pedestrian.}
\label{fig:objects}
\end{figure}

The data of three targets were gathered. Two of them, automated guided vehicle (AGV) and pedestrian, in the indoor scenario, and one, the full size car (2018 black Nissan Micra), in the outdoor scenario. Targets are shown in Fig.~\ref{fig:agv}, Fig.~\ref{fig:uh}, and Fig.~\ref{fig:outdoor1}. The details of targets' sizes and measurement conditions are given in Table~\ref{tb:meas_obj}.

\begin{table}[ht]
	\centering
	\caption{Characteristics of the measured targets.}
	\label{tb:meas_obj}
	\renewcommand{\arraystretch}{1.19}
    \resizebox{0.35\textwidth}{!}{
	\begin{tabular}{|c|c|c|c|}
		\hline
        Dimension  & AGV & Pedestrian & Car \\
        \hline
        Height  & 0.416 m & 1.73 m & 1.455 m \\
        \hline
        Width  & 0.614 m & - & 1.734 m  \\
        \hline
		Length  & 0.72 m & - & 3.99 m  \\
		\hline
        Environment  & Indoor & Indoor & Outdoor  \\
		\hline
        Center & 12 GHz & 12 GHz  & 12.5 GHz  \\
        frequencies  & 28 GHz & 28 GHz &  28 GHz  \\
		\hline
	\end{tabular}}	
\end{table}

\subsection{Indoor environment}\label{IE}
\begin{figure}[htbp]
\centering
\includegraphics[width=0.43\textwidth]{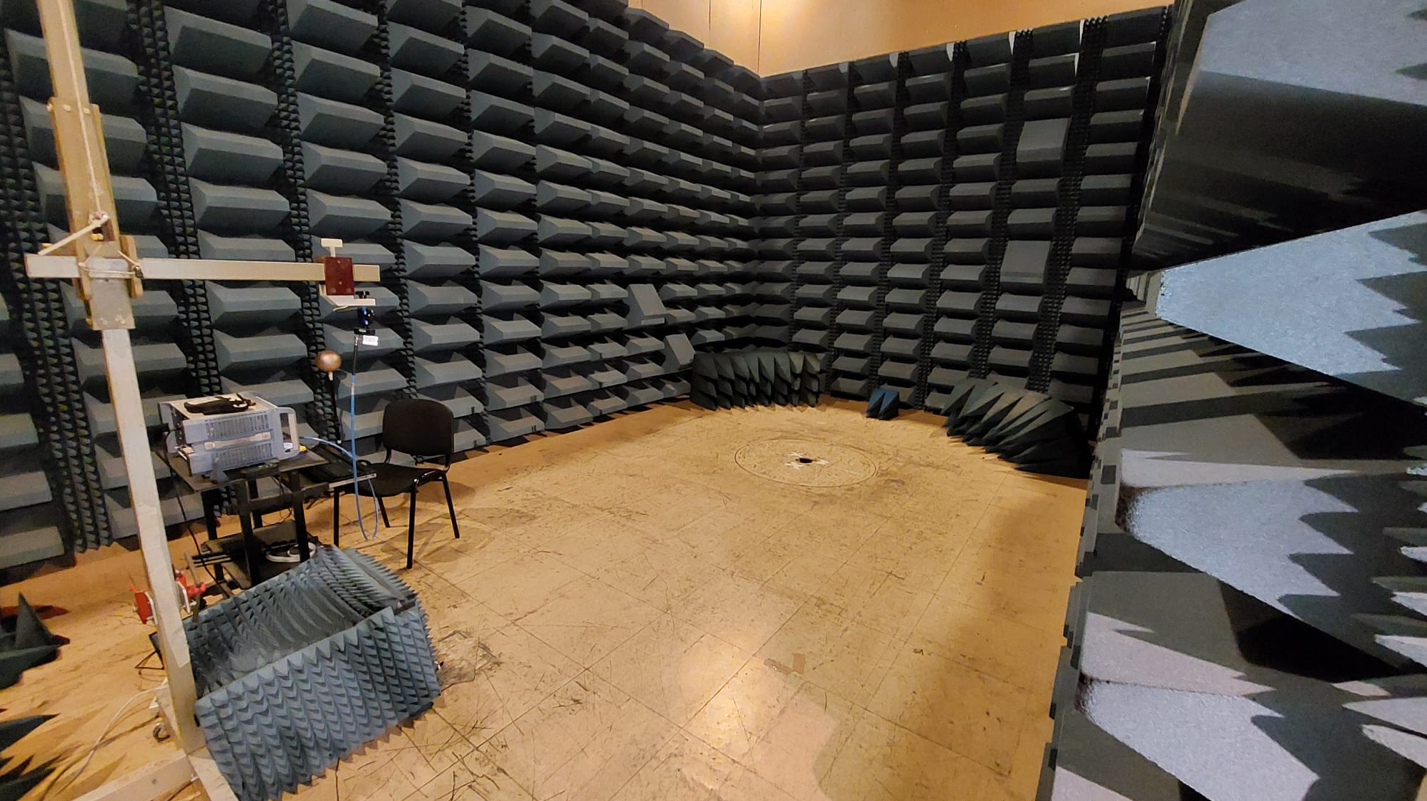}
\caption{Photo of the indoor environment without the dispersive element.}
\label{fig:camara_ph}
\end{figure}

The anechoic chamber used is 4.96~m wide, 8.33~m long, and 6.1~m high, designed for up to 10~GHz use. However, auxiliary absorbing materials of proven efficiency up to 40~GHz were used. Each target has been measured with the antenna at a specified distance so that it is fully covered by the antenna footprint \cite{gowdu2019}. For those measurements whose distance to the target allowed it, a dispersive element has been included to avoid the influence of the reflected component on the ground. Similarly, when it was possible to place the target on the ground, the rotating platform of the anechoic chamber was used. In Fig.~\ref{fig:camara_ph}, a photo of the anechoic chamber is shown.

\vspace{-1pt}

\subsection{Outdoor environment}\label{OE}
\begin{figure}[htbp]
\begin{subfigure}{.24\textwidth}
\includegraphics[width=4.1cm]{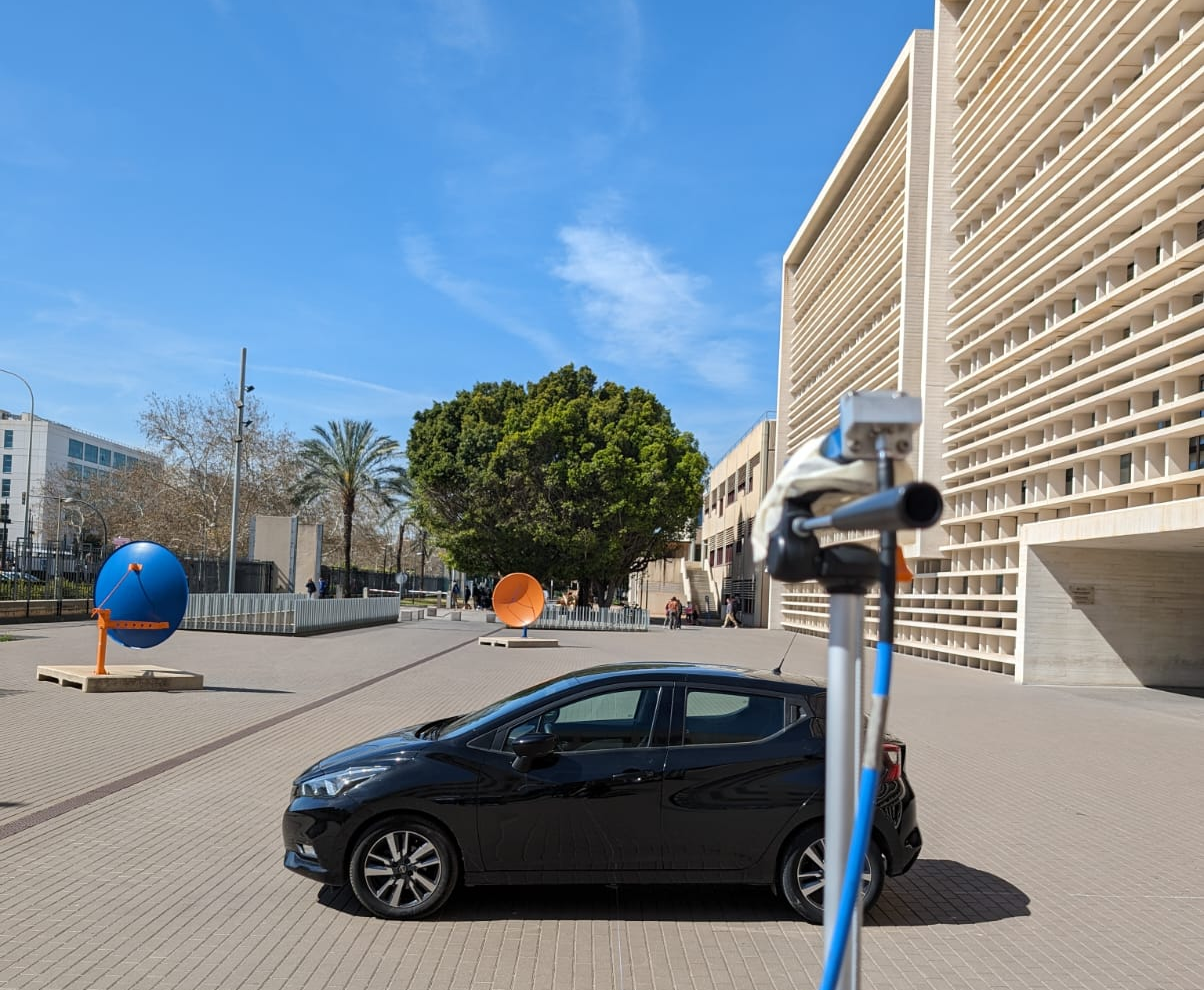}
\caption{}
\label{fig:outdoor1}
\end{subfigure}%
\begin{subfigure}{.24\textwidth}
\includegraphics[width=4.3cm]{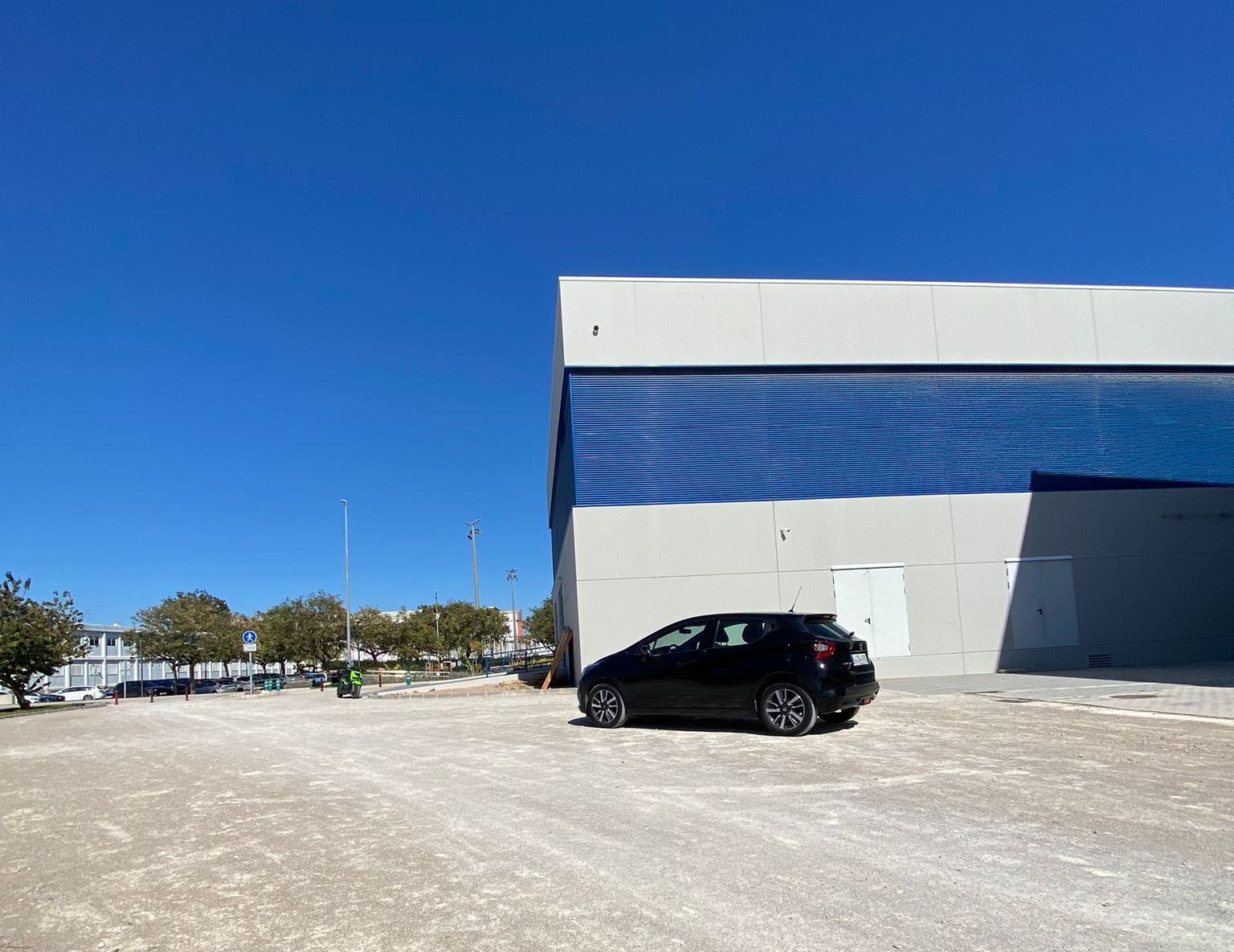}
\caption{}
\label{fig:outdoor2}
\end{subfigure}%
\caption{Photo of the outdoor environment: a) School of Telecommunications Engineering building, b) Research Institute CMT building.}
\label{fig:outdoor}
\end{figure}

Two outdoor scenarios were utilised for the real car measurements. One of them is located in front of the School of Telecommunications Engineering building (Fig.~\ref{fig:outdoor1}). It is an open space with some far-placed scatterers on two sides and buildings on two sides. The second site is located behind the Research Institute CMT building (Fig.~\ref{fig:outdoor2}). Two sides of this area are the building walls; two others are open spaces with distanced greenhouses.
 As in the indoor scenario, it has been designed so that the antenna's footprint covers the entire car. 



\subsection{Measurement procedure}\label{MP}
In order to measure every target properly, two conditions need to be met. The first condition is that the target must be placed in the far field zone according to equation~\ref{eq:farfield}, where $D$ refers to the largest dimension of the antenna's aperture.

\begin{equation}\label{eq:farfield}
    d_{min}=\frac{2\cdot D^{2}}{\lambda }
\end{equation}

\begin{figure}[htbp]
\centering
\includegraphics[width=0.4\textwidth]{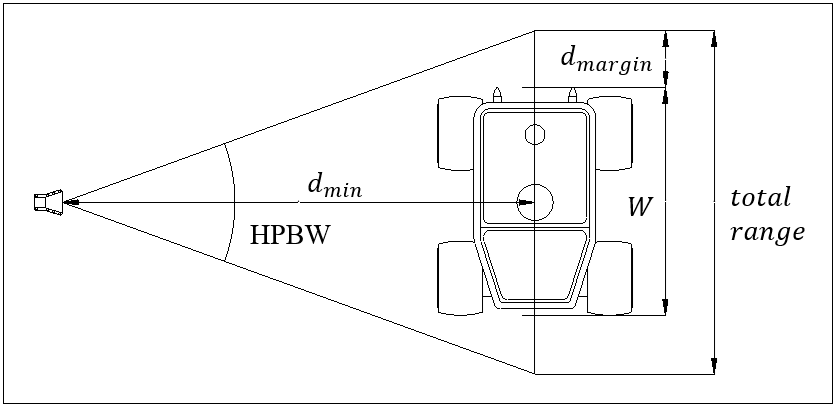}
\caption{Methodology for obtaining the distance related to the antenna footprint.}
\label{fig:azimuthMetho}
\end{figure}

The second condition is related to the antenna footprint on the target under test. In order to capture the entire target inside the antenna footprint, the distance between the antenna and the target must be obtained from the equation \ref{eq:d_min}.
\begin{equation}\label{eq:d_min}
    d_{min} = \biggl( \frac{W}{2}+d_{margin} \biggr) \cdot tan^{-1} \biggl( \frac{HPBW}{2} \biggr),
\end{equation}
where $W$ refers to the width of the target in meters, $d_{margin}$ is a small distance added to avoid approximation errors and $HPBW$ is the half-power beamwidth of the antenna as shown in Fig.~\ref{fig:azimuthMetho}. As every antenna used for this measurement campaign is a single horn antenna, for the working frequency band the far field distance is always less than the minimum distance for the HPBW fitting criteria, so the far field condition is always met for every target. 

\vspace{-2pt}

\vspace{-6pt}

\begin{figure}
  \centering
  \includegraphics[width=0.35\textwidth]{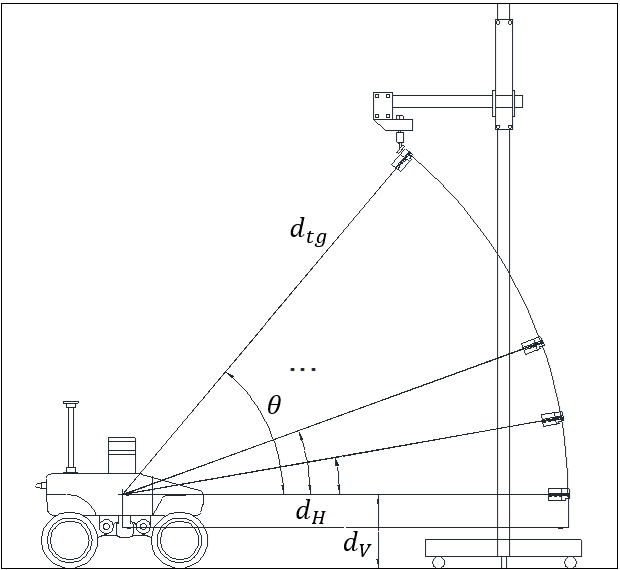}
\caption{Methodology for obtaining the distance between target and antenna related to the antenna elevation.}
\label{fig:elevationhMetho}
\end{figure}

In the case of elevation $\theta$, antennas must be mounted on a tripod, whose height can be adjusted. The methodology employed to reach certain $\theta$ is putting the antenna higher and closer, maintaining the minimum distance as shown in Fig.~\ref{fig:elevationhMetho}. The AGV measurements were conducted for six elevation angles $\theta$ of the antenna: 0$^\circ$, 10$^\circ$, 20$^\circ$, 30$^\circ$, 40$^\circ$ and 50$^\circ$, the pedestrian for 0$^\circ$, 10$^\circ$, and 20$^\circ$, and the car at two $\theta$, 0$^\circ$ and 10$^\circ$.

The methodology followed for the azimuth $\phi$ measurements depends on the scenario and the target; some are above a rotatory platform, others are at a fixed angle, while the antenna is rotating around the target.

In the indoor scenario, a rotatory platform was used for the pedestrian and for the AGV when $\theta>20^\circ$. For angles 0$^\circ$, 10$^\circ$, and 20$^\circ$, AGV was placed on a table and rotated using a remote controller. Targets were turned clockwise with a 10-degree interval. The AGV is symmetrical; thus, when $\theta \leq 20^\circ$ it was measured from 0$^\circ$ to 180$^\circ$. In all other cases, the measurements in indoor scenarios were performed from 0$^\circ$ to 360$^\circ$.

For the outdoor measurement campaign, the target was static, and the antenna was moved around it along a semicircle. The car was measured from 0$^\circ$ to 180$^\circ$ at 10-degree intervals, assuming the symmetry of the vehicle.

\section{Results}


This section presents the RCS obtained from measurements described in Section III. To perform the comparative analysis between frequency bands for each target, a stochastic approximation based on the logarithmic scale differences of the RCS values is used. Finally, the RCS spread in the range domain in both bands is compared.

\subsection{RCS}


The RCS values for the different targets were derived following the methodology outlined in Section II. The adopted representation for displaying the RCS values is a heatmap, where the x-axis denotes the azimuth angle $\phi \degree$ and the y-axis represents the elevation angle $\theta \degree$, corresponding to performed measurements. 

\begin{figure}[htbp]
    \centering
    \begin{subfigure}{0.4\textwidth}
        \centering
        \includegraphics[width=\linewidth]{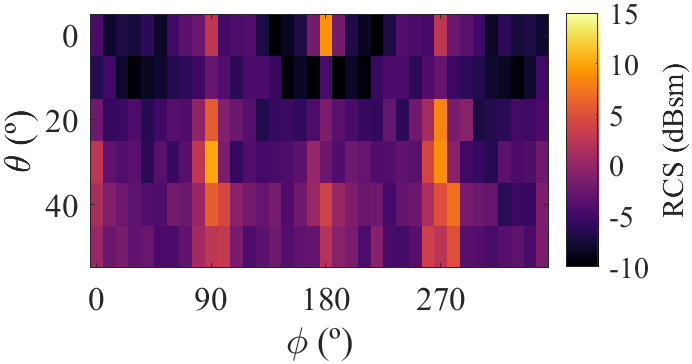}
        \caption{}
        \label{fig:RCS_AGV_midband}
    \end{subfigure}%
    \\
    \begin{subfigure}{0.4\textwidth}
        \centering
        \includegraphics[width=\linewidth]{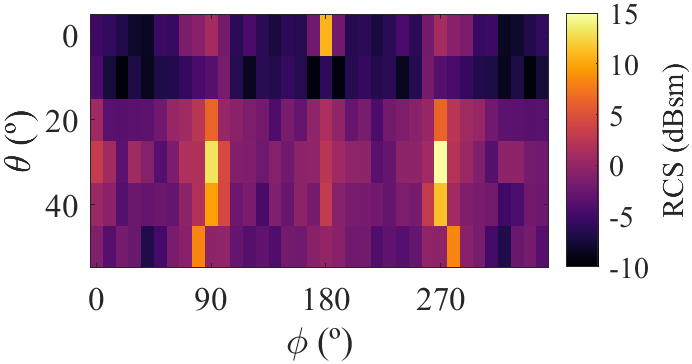}
        \caption{}
        \label{fig:RCS_AGV_mmwave}
    \end{subfigure}%
    \vspace{-4pt}
    \caption{AGV RCS measured for a) $f\in [10, 14] \text{ GHz}$ and b) $f\in [25.75, 30.25] \text{ GHz}$.}
    \label{fig:RCS_AGV}
    \vspace{-15pt}
\end{figure}

In Fig. \ref{fig:RCS_AGV}, the RCS values obtained for the AGV are displayed. The results indicate a notable similarity between the bands in terms of the shape. Bright orange bars, indicating high RCS values, are observed for $\theta=0^\circ$ and $\phi=180^\circ$, and for $\theta$ between 20$^\circ$ and 50$^\circ$, $\phi=90^\circ$ and $\phi=270^\circ$ in both bands. The visual comparison of Fig. \ref{fig:RCS_AGV_midband} and \ref{fig:RCS_AGV_mmwave} indicates the higher of two measured bands gives slightly higher RCS values as the heatmap is brighter. It is also worth noting that the line corresponding to $\theta=10^\circ$ is visibly darker than others.

\vspace{-3pt}

\begin{figure}[htbp]
    \centering
    \begin{subfigure}{0.4\textwidth}
        \centering
        \includegraphics[width=\linewidth]{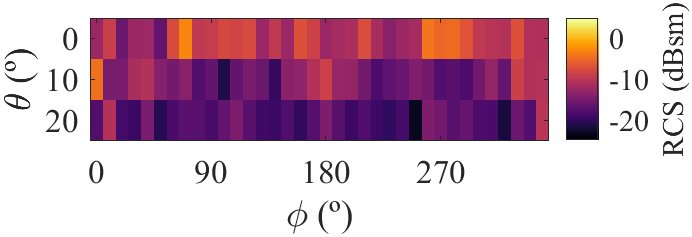}
        \caption{}
        \label{fig:RCS_Human_midband}
    \end{subfigure}%
    \\
    \begin{subfigure}{0.4\textwidth}
        \centering
        \includegraphics[width=\linewidth]{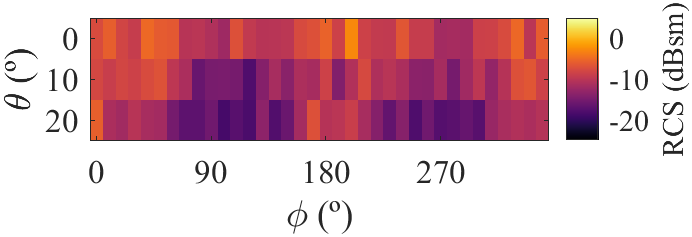}
        \caption{}
        \label{fig:RCS_Human_mmwave}
    \end{subfigure}%
    \vspace{-4pt}
    \caption{Pedestrian RCS measured for a) $f\in [10, 14] \text{ GHz}$ and b) $f\in [25.75, 30.25] \text{ GHz}$.}
    \label{fig:RCS_Pedestrian}
\end{figure}

\begin{figure}[htbp]
    \centering
    \begin{subfigure}{0.4\textwidth}
        \centering
        \includegraphics[width=\linewidth]{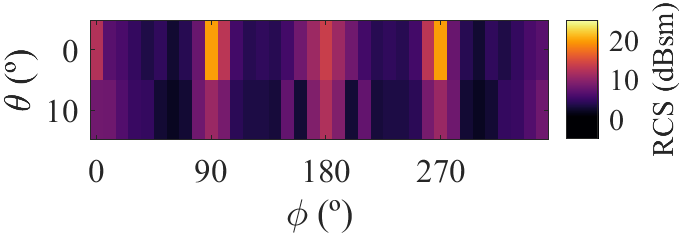}
        \caption{}
        \label{fig:RCS_Car_midband}
    \end{subfigure}%
    \\
    \begin{subfigure}{0.4\textwidth}
        \centering
        \includegraphics[width=\linewidth]{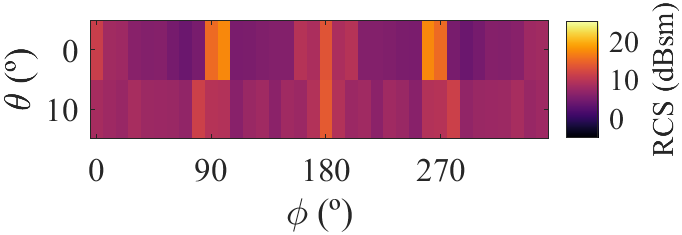}
        \caption{}
        \label{fig:RCS_Car_mmwave}
    \end{subfigure}%
    \caption{Car RCS measured for a) $f\in [10, 15] \text{ GHz}$ and b) $f\in [25.75, 30.25] \text{ GHz}$.}
    \label{fig:RCS_Car}
\end{figure}

For the pedestrian, the RCS results are presented in Fig. \ref{fig:RCS_Pedestrian}. While there is some resemblance in the overall form, the RCS values differ considerably between the frequency bands. It is not possible to visually find the similarities between two frequency bands. However, it can be observed that the highest RCS values are obtained for $\theta=0^\circ$ and the lowest for $\theta=20^\circ$.

Lastly, in Fig. \ref{fig:RCS_Car} the RCS for the car is illustrated. The comparison across bands reveals a pattern analogous to the AGV, with shapes remaining consistent across both frequency bands. Meanwhile, RCS values for the higher bands are noticeably higher.

\subsection{Stochastic approximation to frequency scaling}


Based on the obtained RCS values for both frequency bands and each target, a comparative analysis is performed based on logarithmic scale differences. This analysis is predicated on evaluating RCS behaviour at varying frequencies, aiming to facilitate the scaling of target sizes when utilizing higher frequency bands than the measured one. The RCS is subject to change due to target size scaling and frequency band alterations, making it essential to examine the impact of frequency alterations on RCS. This ensures that any influence due to the used frequency band can be accurately compensated for either through direct target scaling or during post-processing.

\begin{figure*}
    \centering
    \begin{subfigure}{0.3\textwidth}
        \includegraphics[width=\linewidth]{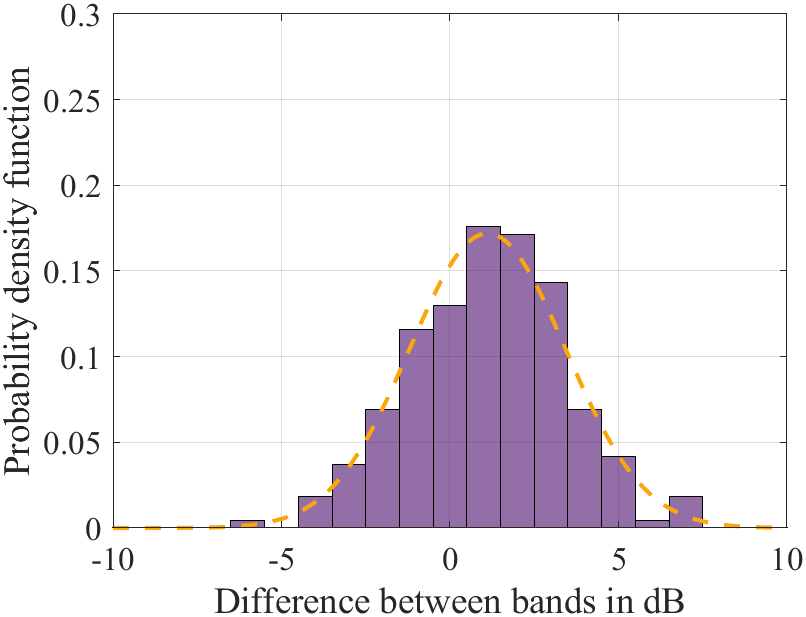}
        \caption{}
        \label{fig:AGV_pdf}
    \end{subfigure}%
    \hfill
    \begin{subfigure}{0.3\textwidth}
        \includegraphics[width=\linewidth]{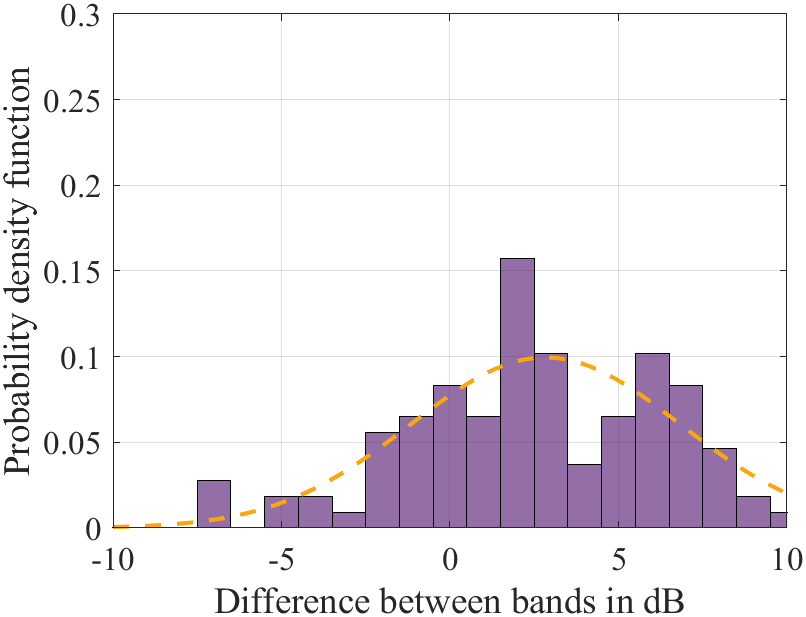}
        \caption{}
        \label{fig:Human_pdf}
    \end{subfigure}%
    \hfill
    \begin{subfigure}{0.3\textwidth}
        \includegraphics[width=\linewidth]{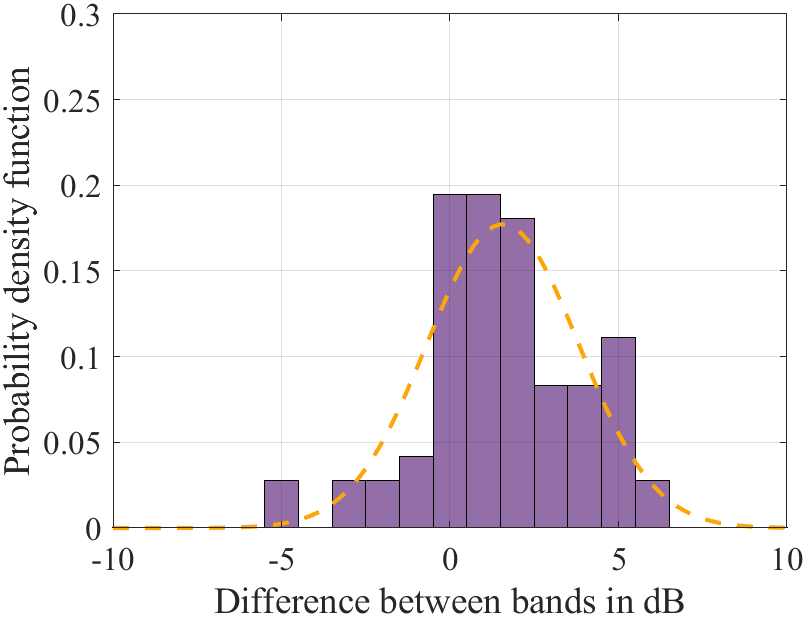}
        \caption{}
        \label{fig:Car_pdf}
    \end{subfigure}
    \caption{Overall $\Delta$RCS histograms for different targets: (a) AGV, (b) Pedestrian, (c) Car.}
    \label{fig:PDF_Targets}
\end{figure*}

As previously mentioned, the metric utilized is the logarithmic scale difference of the RCS values, defined as:

\begin{equation}
    \Delta \text{RCS} = 10\log_{10}\left(\frac{\text{RCS}_{\text{Band 2}}}{\text{RCS}_{\text{Band 1}}}\right),
\end{equation}
where Band 1 corresponds to $f\in[10, 15]$ GHz and Band 2 corresponds to $f\in[25.75, 30.25]$ GHz. For each target, set of $\Delta\text{RCS}$ were calculated. To observe the behaviour of the change between bands, the Gaussian distributions were fitted for each elevation individually, and for all the results obtained for the target. For the overall case, the differences are plotted in histogram format together with their fitted Gaussian curve in Fig. \ref{fig:PDF_Targets}. The mean $\mu$ and standard deviation $\sigma$ results can be found in Table III for every target and every $\theta\degree$. 

\begin{table}[h]
    \centering
    \caption{Gaussian adjustment parameters for $\Delta$RCS in different targets.}
    \resizebox{0.489\textwidth}{!}{
    \begin{tabular}{|c|c|c|c|c|c|c|c|c|}
        \hline
        Target & Parameter & 0º & 10º & 20º & 30º & 40º & 50º & Overall \\
        \hline
        AGV & $\mu$ (dB) & 0.63 & 0.12 & 2.55 & 2.82 & 0.51 & 0.06 & 1.12 \\
            & $\sigma$ (dB) & 1.66 & 1.98 & 1.74 & 2.22 & 2.08 & 2.50 & 2.32 \\
        \hline
        Pedestrian & $\mu$ (dB) & 1.41 & 3.15 & 3.94 & - & - & - & 2.84 \\
                   & $\sigma$ (dB) & 4.37 & 3.11 & 4.11 & - & - & - & 4.01 \\
        \hline
        Car & $\mu$ (dB) & 0.61 & 2.52 & - & - & - & - & 1.57 \\
             & $\sigma$ (dB) & 2.10 & 2.00 & - & - & - & - & 2.25 \\
        \hline
    \end{tabular}}
    \label{tab:combined_targets}
\end{table}

The results indicate that for the AGV, the RCS exhibits minimal variation across frequencies, with average differences around 0-1 dB. Notably, a slight increase in scaling is observed for $\theta = 20^\circ$ and $\theta = 30^\circ$. In the case of the pedestrian, there are considerable differences between the bands, with values nearly 3 dB higher in the $f \in [25.75, 30.25] \, \text{GHz}$ band. The complex surface and geometry of the pedestrian could have affected the longer wavelength band, allowing for new reflections at higher frequencies that increase the RCS. Finally, for the car, the RCS is observed to be higher in the upper band. However, the differences are not comparable to the scaling effects seen in simpler metallic objects like a metal plate or a metal cylinder, whose scalings for the studied frequencies would be approximately 7 dB and 3.5 dB, respectively.

\vspace{-2pt}
\subsection{RCS time-range response}
\vspace{-2pt}

\begin{figure*}[h!]
    \centering
    \begin{subfigure}[b]{0.32\textwidth}
        \centering
        \includegraphics[width=\textwidth]{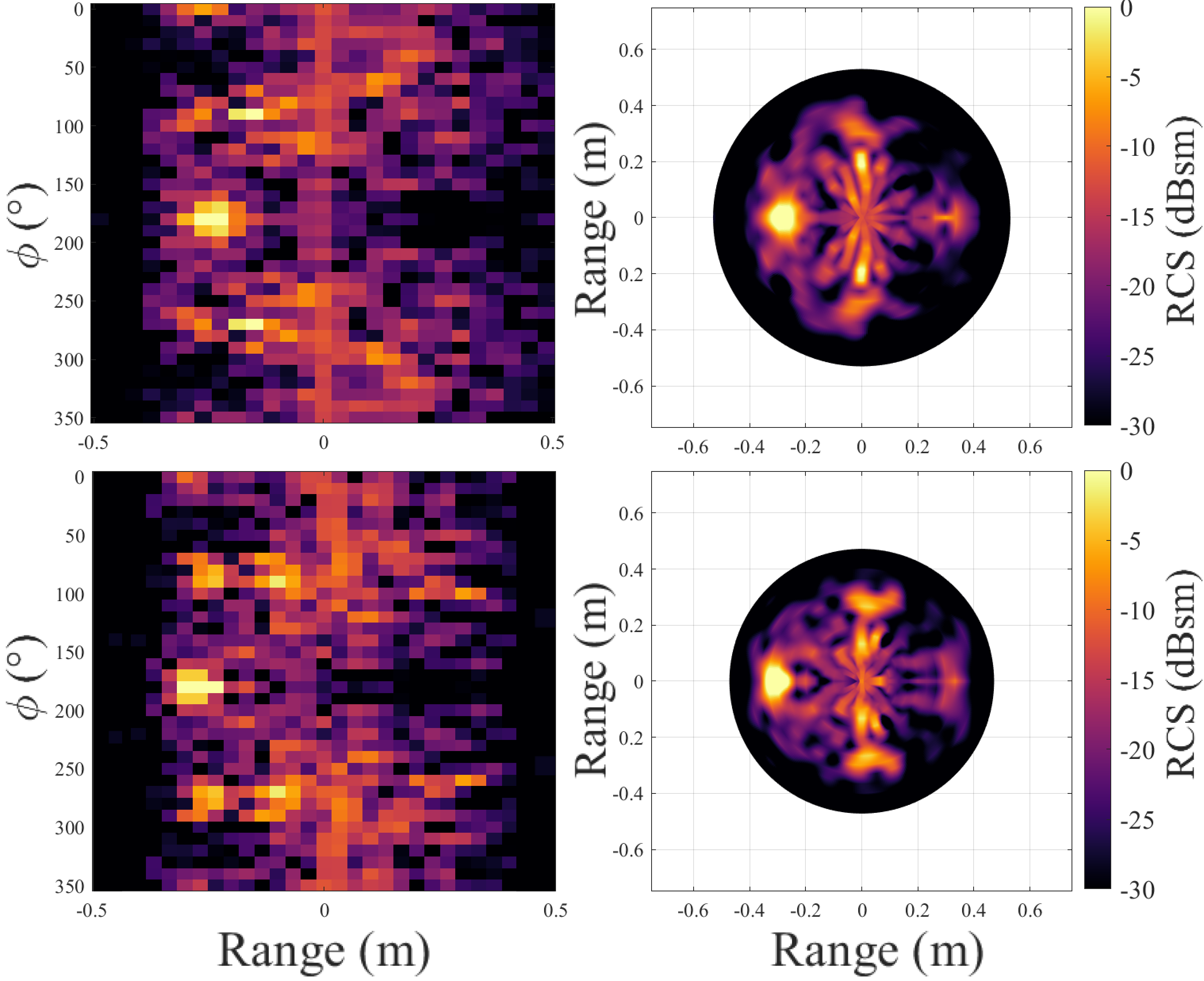}
        \caption{}
        \label{fig:range_response_agv}
    \end{subfigure}
    \hfill
    \begin{subfigure}[b]{0.32\textwidth}
        \centering
        \includegraphics[width=\textwidth]{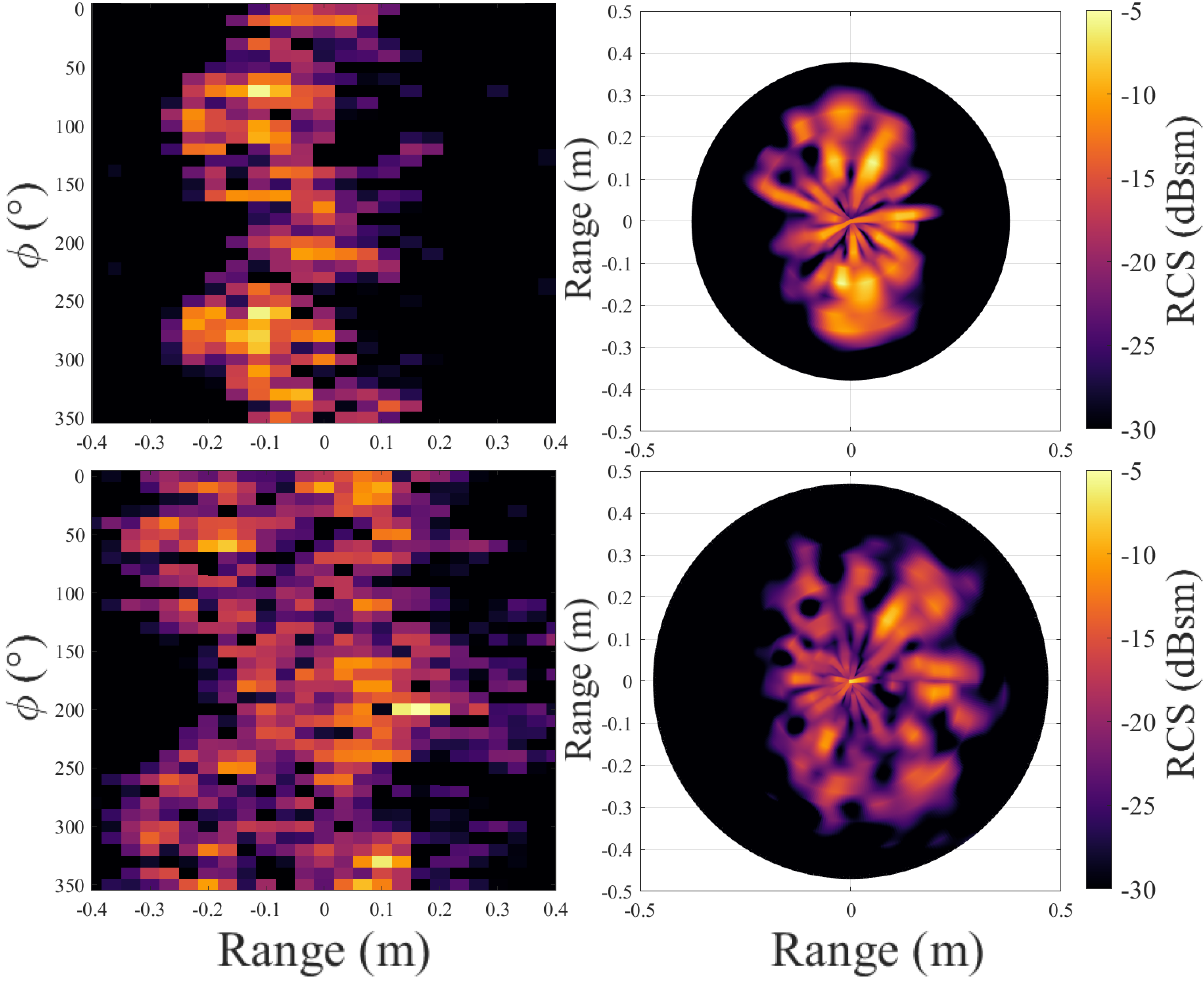}
        \caption{}
        \label{fig:range_response_human}
    \end{subfigure}
    \hfill
    \begin{subfigure}[b]{0.32\textwidth}
        \centering
        \includegraphics[width=\textwidth]{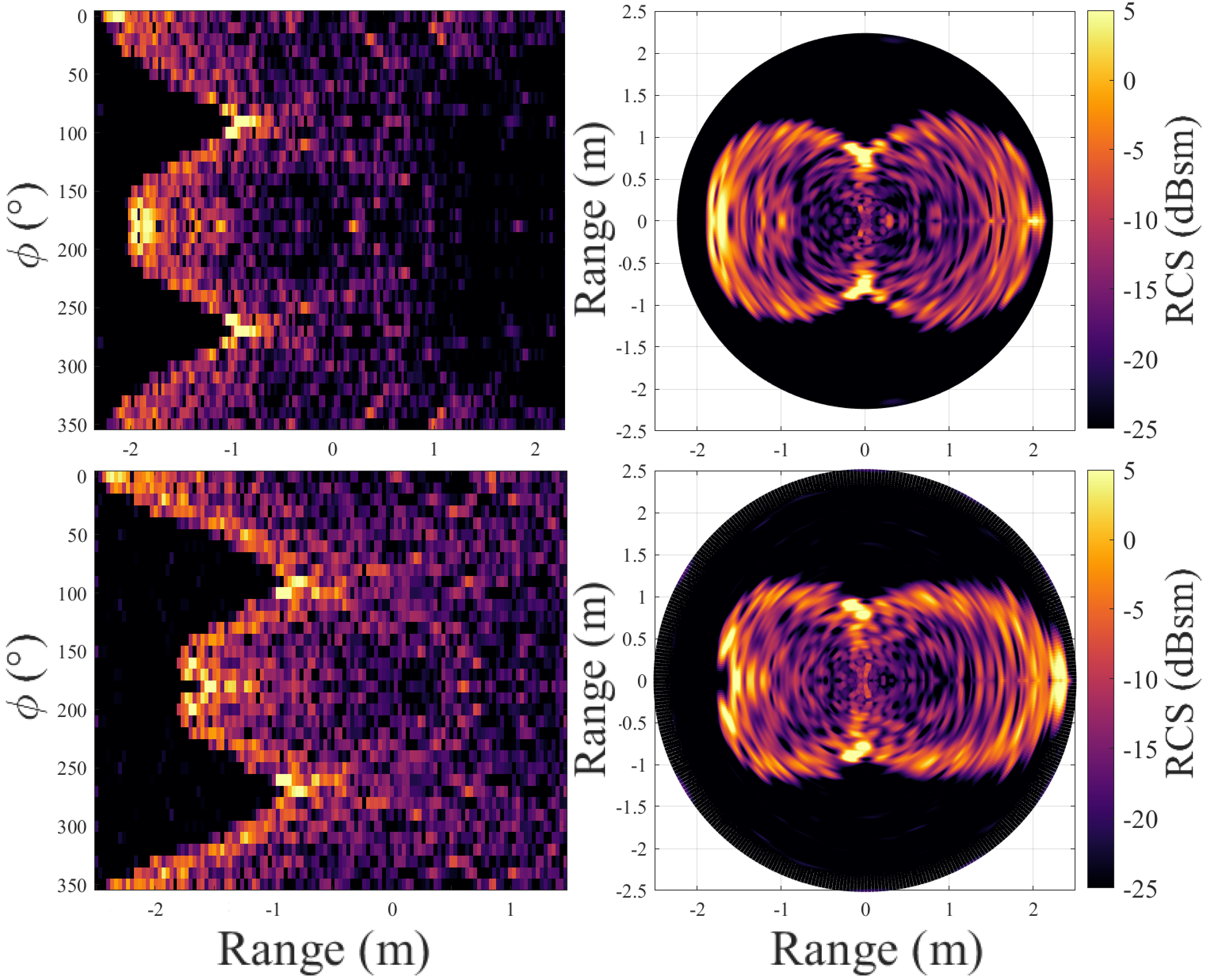}
        \caption{}
        \label{fig:range_response_car}
    \end{subfigure}
    \caption{RCS range response for different objects in $\theta=0\degree$. Upper graphs correspond to $f \in [10, 14\text{-}15] \text{ GHz}$ and lower graphs correspond to $f \in [25.75, 30.25] \text{ GHz}$. (a) AGV, (b) Pedestrian, (c) Car.}
    \label{fig:range_responses}
    \vspace{-9pt}
\end{figure*}

In Section II, $\sqrt{\sigma_{tg}}(t)$ is defined as the square root of the RCS in the time domain. This metric facilitates the assessment of the scattering properties of each target across both frequency bands. Moreover, it provides an empirical method for determining scattering centres (SC), which can be used to simplify complex targets \cite{RCS_SC_1},\cite{RCS_SC_2}. In this analysis, the $\sqrt{\sigma_{tg}}(t)$ in range domain shifted on the target's centre will be compared for both frequencies. This approach allows us to identify the spatial points on the target with the greatest contribution to the RCS, which could potentially serve as precursors to SC. For the comparison between bands, two distinct graphs will be employed for each target (Fig. \ref{fig:range_responses}), all corresponding to $\theta = 0^\degree$. The first graph plots the azimuth angle $\phi$ vs the range, while the second graph represents the range response by positioning each measurement at its corresponding $\phi$, with interpolated values for the unmeasured angles.

The results indicate that for both the AGV and the car (Fig. \ref{fig:range_response_agv} and Fig. \ref{fig:range_response_car}), the distribution of the most influential points across both bands is nearly identical, demonstrating high consistency with respect to frequency variations. However, significant differences are observed in the case of the pedestrian (Fig. \ref{fig:range_response_human}). While the RCS for the lower band is concentrated at specific points, it becomes more spatially distributed when shifting to the higher band.

\section{Conclusions}
\vspace{-2pt}
The conclusions drawn from the study presented in this paper can be summarized in two key aspects: 1) the scaling of RCS values and 2) the spatial points with the greatest influence on them. Firstly, regarding the scaling, it has been observed that variations in frequency lead to variations in RCS values, which depend on the material and shape of the target. For complex shapes and non-metallic materials, such as the pedestrian, the RCS is highly sensitive to frequency. Conversely, for simpler shapes and metallic materials, a certain frequency scaling exists, whose value can be estimated stochastically and must be considered when working with size and frequency scaling.

Regarding the distribution of spatial points with the greatest influence on the RCS, a similar trend has been observed. The pedestrian exhibits a considerably more spatially distributed RCS at higher frequencies, so not being feasibble in this case a generalized frequency characterization based on scattering centers (SC). However, the high similarity in the distribution of points for both the AGV and the car suggests the possibility of developing SC models that are robust to frequency variation, thereby simplifying the modelling of such targets and facilitating their implementation in ISAC channel models.

\vspace{12pt}


\begin{thebibliography}{00}

\bibitem{ISAC_general} F. Liu et al., "Integrated Sensing and Communications: Toward Dual-Functional Wireless Networks for 6G and Beyond," in IEEE Journal on Selected Areas in Communications, vol. 40, no. 6, pp. 1728-1767, June 2022.

\bibitem{ISAC_applications} Y. Cui, F. Liu, X. Jing and J. Mu, "Integrating Sensing and Communications for Ubiquitous IoT: Applications, Trends, and Challenges," in IEEE Network, vol. 35, no. 5, pp. 158-167, September/October 2021.

\bibitem{ISAC_channel1} Z. Zhang et al., "A General Channel Model for Integrated Sensing and Communication Scenarios," in IEEE Communications Magazine, vol. 61, no. 5, pp. 68-74, May 2023.

\bibitem{matsunami2012} I. Matsunami, R. Nakamura and A. Kajiwara, ``RCS measurements for vehicles and pedestrian at 26 and 79GHz," in 2012 6th International Conference on Signal Processing and Communication Systems, (pp. 1-4), IEEE.

\bibitem{schipper2011} T. Schipper, J. Fortuny-Guasch, D., Tarchi, L. Reichardt, and T. Zwick, T, ``RCS measurement results for automotive related objects at 23-27 GHz," in 2011 Proceedings of the 5th European Conference on Antennas and Propagation (EUCAP) (pp. 683-686). IEEE.

\bibitem{ISAC_channel2} Y. Chen, Z. Yu, J. He, J. Li and G. Wang, "A Scatterer-based Hybrid Channel Model for Integrated Sensing and Communications (ISAC)," 2023 IEEE 34th Annual International Symposium on Personal, Indoor and Mobile Radio Communications (PIMRC), Toronto, ON, Canada, 2023, pp. 1-7.

\bibitem{RCS_proc1} W. Hofmann, C. Bornkessel, A. Schwind and M. A. Hein, "Challenges of RF Absorber Characterization: Comparison Between RCS- and NRL-Arch-Methods," in 2019 International Symposium on Electromagnetic Compatibility - EMC EUROPE, Barcelona.

\bibitem{RCS_proc2} R. E. Jarvis, J. G. Metcalf, J. E. Ruyle and J. W. McDaniel, "High Temporal Resolution Time-Gating for Wideband Radar Cross Section Measurements," 2021 51st European Microwave Conference (EuMC), London, United Kingdom, 2022, pp. 998-1001.

\bibitem{ZNB40} Rohde\&Schwarz, "R\&S ZNB/ZNBT Vector Network Analyzers User Manual" [online] Available: https://assets.testequity.com/te1/Documents/pdf/ manuals/znb-m.pdf.

\bibitem{QMS00195} Steatite, "Linearly Polarised Horn Antenna; QSH-SL-10-15-N-20" [online] Available: https://www.steatite-antennas.co.uk/wp-content/uploads/2019/01/QSH-SL-10-15-N-20.pdf.

\bibitem{PE9} Pasternack, "WR-75 Waveguide Standard Gain Horn Antenna" [online] Available: https://www.pasternack.com/wr-75-waveguide-standard-gain-horn-antenna-15-db-sma-pe9855b-sf-15-p.aspx.

\bibitem{QWH1840} Steatite, "Wideband Horn Antenna; QWH-SL-18-40-K-SG" [online] Available: https://qparusa.com/wp-content/uploads/2020/06/QMS-00910\_TEST\_REPORT.pdf.

\bibitem{KBL} Mini-Circuits, "KBL-2M-LOW+" [online] Available: https://www.minicircuits.com/pdfs/KBL-2M-LOW+.pdf.

\bibitem{gowdu2019} S. B. J. Gowdu, A. Schwind, R. Stephan and M. A. Hein, ``Monostatic RCS Measurements of a Passenger Car Mock-up at 77 GHz Frequency in Virtual Environment,'' in 2019 49th European Microwave Conference (EuMC) (pp. 996-999). IEEE.

\bibitem{RCS_SC_1} S. Abadpour, S. Marahrens, M. Pauli, J. Siska, N. Pohl and T. Zwick, "Backscattering Behavior of Vulnerable Road Users Based on High-Resolution RCS Measurements," in IEEE Transactions on Microwave Theory and Techniques, vol. 70, no. 3, pp. 1582-1593, March 2022.

\bibitem{RCS_SC_2} W. Yang et al., "Integrated Sensing and Communication Channel Modeling and Measurements: Requirements and Methodologies Toward 6G Standardization," in IEEE Vehicular Technology Magazine, vol. 19, no. 2, pp. 22-30, June 2024.


\end{thebibliography}
\end{document}